\documentclass[twocolumn,nofootinbib]{revtex4}

\usepackage{blindtext,mathtools,amsmath,amsfonts,amssymb,physics,slashed,esvect,mathrsfs,dsfont,bbold}
\usepackage[utf8]{inputenc}
\usepackage{graphicx}
\usepackage{float}
\usepackage[footnotesize]{caption}

\usepackage{subfig}

\usepackage{multirow}

\usepackage{empheq}

\usepackage{pstricks,pst-node}

\DeclareCaptionJustification{justified}{\leftskip=0pt \rightskip=0pt \parfillskip=0pt plus 1fil}

\usepackage{xcolor}

\mathchardef\Re="023C 
\mathchardef\Im="023D

\pdfoutput=1

\begin{document}
\title{Dark energy explained by an inadequate fitting of the FLRW metric}

\author{Vincent Deledicque}
\email[]{vincent.deledicq@gmail.com}
\affiliation{No affiliation}

\date{\today}

\begin{abstract}
Approximating a real manifold by an idealized one requires to calibrate the parameters characterizing the idealized manifold in function of the real one. This calibration is a purely conventional process and can generally be done in several ways, leading to different fittings. In practice, however, all possible fittings cannot be considered as representative of the real manifold. Approximating the real metric of the universe by the FLRW metric would be adequate only if both corresponding structures, defined by the space-time interval, are equivalent on large scales. This requirement puts some constraints on what would be a representative FLRW metric. We show that the way how measurements on SNIa are interpreted to determine the evolution of the scale factor implicitly define the calibration process, and that this one is compatible with the aforementioned constraints. On a theoretical point of view, this indicates that the as fitted FLRW metric would indeed be representative of the real one. On a practical point of view, however, we show that a bias in the measurements could invalidate this conclusion. The bias comes from the fact that SNIa are not randomly distributed over space, but are probably mostly located in regions were matter is largely present, i.e., in overdense regions. We explain how this bias could account for the apparent accelerated expansion of the universe, without needing to introduce the dark energy assumption. We show in particular that this bias leads to an inadequate fitting of the FLRW metric, resulting in the appearance of a new term in the evolution equation of the related scale factor, being equivalent to the cosmological constant. 
\end{abstract}

\maketitle

%----------------------------------------------------------------------------------------

\section{Introduction}

Typical cosmological models are based on the postulate that the universe is homogeneous and isotropic in its spatial dimensions. This postulate is generally known as the Cosmological principle. Obviously, at small scales, the universe presents heterogeneities and anisotropies, but we take the Cosmological principle to apply only on the largest scales, where local variations are averaged over. The homogeneity and isotropy of the universe at such scales imply that space would be maximally symmetric, leading to the well-known Friedmann-Lema\^\i tre-Robertson-Walker (FLRW) metric. In typical cosmological models, this metric is then used to determine the left part of the Einstein equation of general relativity:
\begin{eqnarray} \label{Einsteinequation}
R_{\mu\nu} - \frac{1}{2}g_{\mu\nu}R + \Lambda g_{\mu\nu}  = 8\pi G T_{\mu\nu}\,,
\end{eqnarray}
where $\Lambda$ is the cosmological constant. The right part is determined by estimating the average stress-energy tensor of all identified sources. Solving the Einstein equation leads finally to the Friedmann equations, allowing to predict the behavior of the scale factor $a(t)$ of the universe in function of the energy density and the pressure. In simple words, this approach allows to predict the global evolution of the universe in function of its content. 

It is known since the beginning of the study of cosmology that space is not perfectly homogeneous and isotropic, but the effects of this characteristic on the evolution of the universe have been investigated seriously since some decades only. As highlighted by \cite{Clarkson}, considering heterogeneity and anisotropy involves several major difficulties, related in particular to the fitting problem, to the averaging of the Einstein equations, and to the investigation of possible backreaction effects.

Important efforts have been put by several authors on the investigation of backreaction effects, see \cite{Buchert}, \cite{Clifton}, \cite{Kolb} and many others. In such studies, the accelerating expansion of the universe, as evidenced by \cite{Riess} and \cite{Perlmutter} from the observations of distant Type Ia supernovae (SNIa), is explained by the fact that the Einstein equation of general relativity should not be applied as such for the FLRW metric, but should be averaged in some way to take into account the inhomogeneous reality. The non-commutativity of the averaging procedure would lead to new terms in the averaged Einstein equations that would account for the observed acceleration. Also, many discussions have been held on how to best define this averaging process, see for example \cite{Hoogen} and \cite{Wiltshire}.

While backreaction effects and the difficulties of averaging the Einstein equations have received significant attention, the fitting problem has regrettably received much less attention. This last one is however a fundamental issue, because the fitting process defines the idealized universe we are studying, and which is supposed to be representative of the real universe. An inadequate fitting would imply that all conclusions drawn from the analysis of the idealized universe could not necessarily be transposed to the real one. 	

Before observations led to the conclusion that space-time undergoes an accelerated expansion, \cite{Ellis} studied what would be the best way to fit an idealized homogeneous and isotropic universe to a realistic one. This was an important question, but which was not really taken into consideration when estimating the evolution of the scale factor from the observations on SNIa. However, as we will see, the way observations have been interpreted implicitly define how the idealized universe has been fitted on the real universe. So the question would now become: is this fitting adequate? In other words, may we consider that the behavior of the as fitted FLRW universe is representative of the behavior of the real universe viewed at a global scale? This is a fundamental question, and if the answer is no, all other investigations to explain the observed behavior could be irrelevant.

\begin{figure}
	\centering\includegraphics[width=6cm]{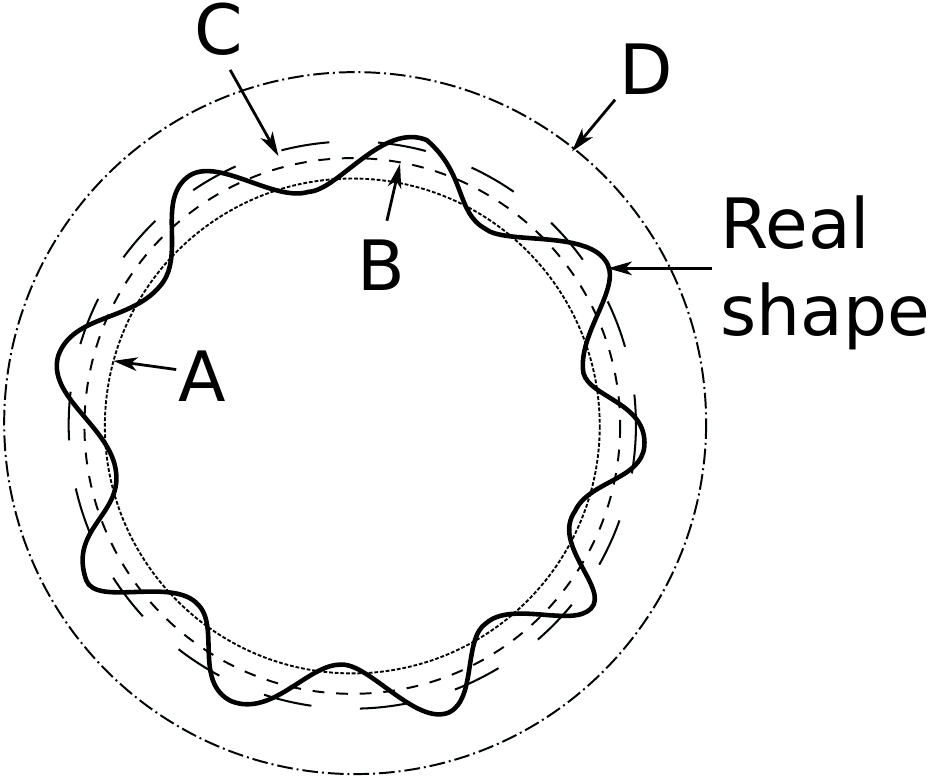}
	\caption{Illustration of different fittings of a circle on an approximative circular shape.}\label{Cercle}
\end{figure}

The importance of this question can be illustrated as follows: when approximating a real manifold by an idealized one, the parameters characterizing this idealized manifold have to be calibrated in some way. This calibration is the mathematical description of the fitting process. It is a conventional process, and it can be done in several ways, meaning that different approximations of the real manifold by the idealized one could be obtained. Let us consider the simple example of some growing shape that looks like a circle, but which presents locally small perturbations with respect to a perfect circle, see Figure $\ref{Cercle}$. For illustration purposes, the perturbations have been willingly disproportionately enlarged on that figure. When looking how this shape evolves over time, it makes sense, for practical reasons, to approximate it as a simple circle and claim that the evolution of this circle is representative of the evolution of the real shape. However, an infinite number of possibilities exists for this approximation, 4 of which are illustrated on Figure $\ref{Cercle}$. It is not difficult to imagine cases for which approximation $A$ would grow with an decreasing rate, approximation $B$ would grow with a constant rate, whereas approximation $C$ would grow with a increasing rate. The important result of this simple illustration is that for a given shape, different fittings of its approximate circle could lead to different conclusions about the way it expands over time. Moreover, some measurements performed on the real shape and inadequately interpreted could even lead to approximation $D$, which could hardly be considered as representative of the real shape. In this simple case, it can easily be seen that such a fitting is inadequate, but this is not always as easy for more complex and multidimensional manifolds.

The same could happen for space-time. Space is not perfectly homogeneous and isotropic, and the universe's metric does not exactly correspond to the FLRW metric. Obviously, since it probably looks like the FLRW metric, it makes sense to approximate it as such, but this approximation can be done in several ways. It is important to know exactly how the approximation is carried out, because this could have an impact on how space-time will be seen evolving. In practice, the approximation is implicitly defined by the way measurements on SNIa are interpreted. The aim of this article is to investigate if this approximation can be considered as representative of the real universe, and if not, to investigate if this inadequate approximation could account for the apparent accelerated expansion of the universe.

In section $\ref{S2}$ we first establish relations between the real metric and the FLRW metric, as well as for derived tensors and scalars, by considering a perturbation approach. Such relations will be useful in the investigation. In section $\ref{S3}$ we define the best way to approximate the real metric by a FLRW metric, and identify the constraints implicitly imposed by this definition. In section $\ref{S4}$ we then use these constraints to determine the evolution laws of the FLRW metric, which are shown to be identical to the Friedmann equations. In section $\ref{S5}$ we finally show that the way we interpret the luminosity distance and redshift measurements implies the same constraints as the ones imposed by the definition of the representative FLRW metric. On a theoretical point of view, this indicates that the as fitted FLRW metric would indeed be representative of the real one. We explain however in section $\ref{S6}$ that a bias in the measurements could invalidate this conclusion. The bias comes from the fact that SNIa are not randomly distributed over space, but are probably mostly located in regions were matter is largely present, i.e. in overdense regions. We explain how this bias could account for the apparent accelerated expansion of the universe, without needing to introduce the dark energy assumption.

%--------------------------------------------------------------------

\section{The perturbation approach}\label{S2}

Besides the analysis of possible backreaction effects, as highlighted by \cite{Clarkson}, two major difficulties arise when analyzing the evolution of space-time by considering that it behaves as having a global average metric. The first difficulty is related to the fitting of the FLRW metric. The second difficulty lies in establishing the average of the Einstein equation, which allows determining the evolution law for the scale factor characterizing this FLRW metric, taking into account how the FLRW metric has been calibrated.

Both difficulties require to relate in some way the real metric to the approximate FLRW metric. Before addressing the two aforementioned difficulties, we will formulate the different scalars and tensors derived from the FLRW metric in function of their equivalent ones derived from the real metric, by using a perturbation method. This is a well-known approach, whose derivation can be found in several textbooks. For convenience, the main expressions of this perturbation approach that will be useful for our investigation will be rewritten here.

We will temporarily admit that we are able to define the average FLRW metric, which will be written as $\overline{g}_{\mu\nu}$. Note that in this article we will only consider the case of a flat FLRW metric. Therefore, expressed in a $(t,x,y,z)$ reference frame, we have $\overline{g}_{tt} = -1$, $\overline{g}_{ii} = a^2(t)$ where $i=x,y,z$, and all other components are zero.

Locally, the real metric will be written as $g_{\mu\nu}$. The difference between the local metric and the average FLRW metric, called the perturbation, is written as $\delta g_{\mu\nu}$. Hence, at each point of space-time, we have
\begin{equation}
g_{\mu\nu} = \overline{g}_{\mu\nu} + \delta g_{\mu\nu}\,.
\end{equation}
Generally, $\delta g_{\mu\nu}$ depends on $x$, $y$, $z$ and $t$, whereas $\overline{g}_{\mu\nu}$ only depends on $t$ at most. The inverse $g^{\alpha\mu}$ is such that $g^{\alpha\mu}g_{\mu\nu} = \delta^\alpha_{\ \nu}$. Similarly, $\overline{g}^{\alpha\nu}$ is the inverse of $\overline{g}_{\mu\nu}$.

We assume that the perturbation $\delta g_{\mu\nu}$ is generally quite small with respect to $g_{\mu\nu}$. It is then easy to convince us that
\begin{equation}\label{gperturb}
g^{\alpha\mu} = \overline{g}^{\alpha\mu} - \delta g^{\alpha\mu}\,.
\end{equation}
For completeness, we also assume that the first and second partial derivatives, $\delta g_{\mu\nu,\rho}$ and $\delta g_{\mu\nu,\rho\sigma}$, are small with respect to $g_{\mu\nu,\rho}$ and $g_{\mu\nu,\rho\sigma}$, respectively. 

A useful relation for our investigation in the next sections can be obtained starting from the following property:
\begin{eqnarray}
\delta g^{\mu\nu} &=& \left(\overline{g}^{\alpha\mu}-\delta g^{\alpha\mu}\right)\left(\overline{g}^{\beta\nu}-\delta g^{\beta\nu}\right)\delta g_{\alpha\beta} \nonumber
\\
&\simeq& \overline{g}^{\alpha\mu}\overline{g}^{\beta\nu}\delta g_{\alpha\beta}\,,
\end{eqnarray}
where in the last equation second order terms have been neglected. Differentiating this equation with respect to $\sigma$, we find that:
\begin{equation}
\delta g^{\mu\nu}_{\ \ \,,\sigma} = \overline{g}^{\alpha\mu}_{\ \ \,,\sigma}\overline{g}^{\beta\nu}\delta g_{\alpha\beta} + \overline{g}^{\alpha\mu}\overline{g}^{\beta\nu}_{\ \ \,,\sigma}\delta g_{\alpha\beta} + \overline{g}^{\alpha\mu}\overline{g}^{\beta\nu}\delta g_{\alpha\beta,\sigma}\,.
\end{equation}

We may now determine the Christoffel symbols in function of the FLRW metric. At each point of space-time we have
\begin{eqnarray}
\Gamma^\alpha_{\ \beta\gamma} &=& \frac{1}{2}\left(\overline{g}^{\alpha\nu} - \delta g^{\alpha\nu}\right)\left(\overline{g}_{\nu\gamma,\beta} + \delta g_{\nu\gamma,\beta} + \overline{g}_{\beta\nu,\gamma}\right.\nonumber
\\
&&\left. + \delta g_{\beta\nu,\gamma} - \overline{g}_{\beta\gamma,\nu} - \delta g_{\beta\gamma,\nu}\right)\,.
\end{eqnarray}
When developing this expression, we get
\begin{equation}
\Gamma^\alpha_{\ \beta\gamma} = \overline{\Gamma}^\alpha_{\ \beta\gamma} + \delta \Gamma^\alpha_{\ \beta\gamma}\,,
\end{equation}
where
\begin{equation}\label{hihi}
\overline{\Gamma}^\alpha_{\ \beta\gamma} = \frac{1}{2}\overline{g}^{\alpha\nu}\left(\overline{g}_{\nu\gamma,\beta} + \overline{g}_{\beta\nu,\gamma} - \overline{g}_{\beta\gamma,\nu}\right)
\end{equation}
is the part of the Christoffel symbols calculated on the basis of the FLRW metric only, and where
\begin{eqnarray}\label{hoho}
\delta \Gamma^\alpha_{\ \beta\gamma} &=& \frac{1}{2}\overline{g}^{\alpha\nu}\left(\delta g_{\nu\gamma,\beta} + \delta g_{\beta\nu,\gamma} - \delta g_{\beta\gamma,\nu}\right)\nonumber
\\ 
&& - \frac{1}{2}\delta g^{\alpha\nu}\left(\overline{g}_{\nu\gamma,\beta} + \delta g_{\nu\gamma,\beta} + \overline{g}_{\beta\nu,\gamma} + \delta g_{\beta\nu,\gamma}\right.\nonumber
\\
&&\left. - \overline{g}_{\beta\gamma,\nu} - \delta g_{\beta\gamma,\nu}\right)
\end{eqnarray}
is the perturbation that has to be added to $\overline{\Gamma}^\alpha_{\ \beta\gamma}$ in order to determine the local Christoffel symbols. Expressed in a $(t,x,y,z)$ reference frame, we have $\overline{\Gamma}^t_{\ ii} = a\dot{a}$, $\overline{\Gamma}^i_{\ it} = \dot{a}/a$, where $i=x,y,z$, and all other components are zero.

Next, the Ricci tensor can be written as
\begin{equation}
R_{\alpha\beta} = \overline{R}_{\alpha\beta} + \delta R_{\alpha\beta}\,,
\end{equation}
where
\begin{equation}
\overline{R}_{\alpha\beta} = \overline{\Gamma}^\mu_{\ \alpha\beta,\mu} - \overline{\Gamma}^\mu_{\ \alpha\mu,\beta} + \overline{\Gamma}^\mu_{\ \alpha\beta}\overline{\Gamma}^\nu_{\ \mu\nu} - \overline{\Gamma}^\nu_{\ \alpha\mu}\overline{\Gamma}^\mu_{\ \nu\beta}
\end{equation}
is the part of the Ricci tensor calculated on the basis of the FLRW metric only, and where
\begin{eqnarray}\label{erpao}
\delta R_{\alpha\beta} &=& \overline{\Gamma}^\mu_{\ \alpha\beta}\delta \Gamma^\nu_{\ \mu\nu} + \overline{\Gamma}^\nu_{\ \mu\nu}\delta \Gamma^\mu_{\ \alpha\beta} + \delta \Gamma^\mu_{\ \alpha\beta}\delta \Gamma^\nu_{\ \mu\nu}\nonumber
\\
&& - \overline{\Gamma}^\nu_{\ \alpha\mu}\delta \Gamma^\mu_{\ \nu\beta} - \overline{\Gamma}^\mu_{\ \nu\beta}\delta \Gamma^\nu_{\ \alpha\mu} - \delta \Gamma^\nu_{\ \alpha\mu}\delta \Gamma^\mu_{\ \nu\beta}\nonumber
\\
&& +\delta \Gamma^\mu_{\ \alpha\beta,\mu} - \delta \Gamma^\mu_{\ \alpha\mu,\beta}
\end{eqnarray}
is the perturbation that has to be added to $\overline{R}_{\alpha\beta}$ in order to determine the local Ricci tensor. Expressed in a $(t,x,y,z)$ reference frame, we have $\overline{R}_{tt} = -3\ddot{a}/a$, $\overline{R}_{ii} = a\ddot{a}+2\dot{a}^2$ where $i=x,y,z$, and all other components are zero.

Finally, we write the Ricci scalar as
\begin{equation}\label{Rperturb}
R = \overline{R} + \delta R\,,
\end{equation}
where
\begin{equation}
\overline{R} = \overline{g}^{\alpha\beta}\overline{R}_{\alpha\beta}
\end{equation}
is the part of the Ricci scalar calculated on the basis of the FLRW metric only, and where
\begin{equation}\label{deltaR}
\delta R = \overline{g}^{\alpha\beta}\delta R_{\alpha\beta} - \overline{R}_{\alpha\beta}\delta g^{\alpha\beta} - \delta g^{\alpha\beta}\delta R_{\alpha\beta}
\end{equation}
is the perturbation that has to be added to $\overline{R}$ in order to determine the local Ricci scalar. For the FLRW metric we have $\overline{R} = 6\left(\ddot{a}/a+\dot{a}^2/a^2\right)$.

At each point of space-time, admitting that the cosmological constant is zero, the Einstein equation reads
\begin{eqnarray}\label{ttt}
R_{\mu\nu} - \frac{1}{2}g_{\mu\nu}R = 8\pi G T_{\mu\nu}\,.
\end{eqnarray}
Using the expressions developed above, this last relation can be written as
\begin{equation}\label{Eq22}
\overline{G}_{\mu\nu} + \delta G_{\mu\nu} = 8\pi G T_{\mu\nu} \,,
\end{equation}
where $\overline{G}_{\mu\nu} = \overline{R}_{\mu\nu} - 1/2\overline{g}_{\mu\nu}\overline{R}$ is the Einstein tensor defined on the basis of the FLRW metric, and where we defined 
\begin{eqnarray}\label{deltaG}
\delta G_{\mu\nu} = \delta R_{\mu\nu} - \frac{1}{2}\left(\overline{g}_{\mu\nu}\delta R + \delta g_{\mu\nu}\overline{R} + \delta g_{\mu\nu}\delta R\right)\,.
\end{eqnarray}
Expressed in a $(t,x,y,z)$ reference frame, we have $\overline{G}_{tt} = 3\dot{a}^2/a^2$, $\overline{G}_{ii} = -2a\ddot{a}-\dot{a}^2$ where $i = x,y,z$, and all other components are zero. 

%--------------------------------------------------------------------

\section{The representative FLRW metric}\label{S3}

The manifold defined by the FLRW metric is such that
\begin{equation}\label{78}
d\overline{s}^2 = \overline{g}_{\mu\nu}dx^\mu dx^\nu\,.
\end{equation}
On the other hand, the manifold defined by the real metric is such that
\begin{equation}\label{79}
ds^2 = \left(\overline{g}_{\mu\nu} + \delta g_{\mu\nu}\right)dx^\mu dx^\nu\,.
\end{equation}
Approximating the real manifold expressed by Eq.\ $(\ref{79})$ by the idealized one expressed by Eq.\ $(\ref{78})$ would be meaningless if, over large scales, both were not equivalent. This equivalence means that on large scales, the integration of Eq.\ $(\ref{79})$ on some path should converge to the integration of Eq.\ $(\ref{78})$ on the same path. If this was not the case, the FLRW metric would not constitute a representative approximation of the real metric, in particular because spatial distances or time intervals would globally be different between the two manifolds. Spatial distances and time intervals are important notions when measuring the time evolution of the scale factor. If distances or time intervals as calculated from the FLRW metric would not converge to the ones as calculated from the real metric on large scales, the approximate manifold (and the temporal evolution of its scale factor) could hardly be considered as representative of the real one.

Let us consider a spatial path that follows the $x$-axis ($dt = dy = dz = 0$), and let us integrate Eq.\ $(\ref{79})$ along that path. This leads to 
\begin{equation}\label{88}
\int_x ds = \int_x \sqrt{g_{xx}}dx = \int_x \sqrt{\overline{g}_{xx} + \delta g_{xx}}dx\,,
\end{equation}
where $\int_x$ means that the integration is performed along the $x$-axis. The limits of the integration are not written for convenience, but it is assumed that the integration path is sufficiently large so that perturbations are averaged over. On large scales, we expect that:
\begin{equation}\label{89}
\int_x \sqrt{g_{xx}}dx \longrightarrow \int_x \sqrt{\overline{g}_{xx}}dx\,.
\end{equation}
On a theoretical point of view, if we knew the metric $g_{\mu\nu}$ at each point, this would directly allow to deduce the scale factor. Indeed, at the limit of Eq.\ $(\ref{89})$ we simply get
\begin{equation}\label{amoy}
a = \frac{\int_x \sqrt{g_{xx}}dx}{\int_x dx}\,.
\end{equation}
Obviously, given the isotropy of space, a same value would be obtained by considering an integration path in a different direction. 

The knowledge of the scale factor completely determines the spatially flat FLRW metric, which in some sense can be considered as the average of the real metric. A lot of discussions have been held about the difficulties related to how tensors (and in particular the metric tensor) should be averaged, see \cite{Hoogen} for example. However, imposing the structure of real space, as defined by the space-time interval, to be identical on large scales as the one determined by the FLRW metric provides a simple definition of this average FLRW metric. It is important to note that this definition is not based on the volume integration of some tensor, which is a questionable operation, but on the integration of the interval $ds$ which is a scalar. It is thus well defined and frame invariant.	

Before going further, let us deduce some interesting consequences from this definition. The different relations that we will establish hereunder will be useful in the next section.

First, still at the limit expressed by Eq.\ $(\ref{89})$, we find that
\begin{equation}\label{zef}
\int_x \left(\sqrt{\overline{g}_{xx} + \delta g_{xx}} - \sqrt{\overline{g}_{xx}} \right)dx = 0\,.
\end{equation}
Using a Taylor expansion to approximate the first square root, we have
\begin{equation}
\sqrt{\overline{g}_{xx} + \delta g_{xx}} \simeq \sqrt{\overline{g}_{xx}} + \frac{\delta g_{xx}}{2\sqrt{\overline{g}_{xx}}}\,,
\end{equation}
and Eq.\ $(\ref{zef})$ simplifies as
\begin{equation}\label{134}
\int_x \frac{\delta g_{xx}}{2\sqrt{\overline{g}_{xx}}}dx = 0\,.
\end{equation}
But since $\overline{g}_{xx} = a^2(t)$ does not depend on $x$, Eq.\ $(\ref{134})$ simply implies
\begin{equation}\label{Cons1}
\int_x \delta g_{xx}dx = 0\,.
\end{equation}
Strictly speaking, this identity will in general not be verified exactly, and it should be considered as an approximation. The meaning of Eq.\ $(\ref{Cons1})$ is that if $\delta g_{xx}$ can sometimes be positive or negative, on average over space it is zero. It is important to stress that this result is a direct consequence of the way we defined the representative FLRW metric. For other definitions, Eq.\ $(\ref{Cons1})$ would not be verified in general.

It is important to highlight that Eq.\ $(\ref{Cons1})$ is valid at each time:
\begin{equation}
\int_x \delta g_{xx}(t)dx = \int_x \delta g_{xx}(t+\delta t)dx = 0\,,
\end{equation}
where for convenience we did only specify the temporal dependence of $\delta g_{xx}$, its dependence on $x$, $y$ and $z$ being implicitly assumed. Since we can approximate 
\begin{equation}
\delta g_{xx}(t+\delta t) \simeq \delta g_{xx}(t) + \delta g_{xx,t}(t)\delta t\,,
\end{equation}
we find that
\begin{equation}\label{Cons111}
\int_x \left[\delta g_{xx}(t) + \delta g_{xx,t}(t)\delta t\right]dx = 0\,.
\end{equation}
Subtracting Eq.\ $(\ref{Cons1})$ from Eq.\ $(\ref{Cons111})$, we deduce that
\begin{equation}
\int_x \delta g_{xx,t}(t)dx = 0\,.
\end{equation}
Since this relation once again holds for all times, we may in a similar way show that 
\begin{equation}
\int_x \delta g_{xx,tt}(t)dx = 0\,.
\end{equation}

Finally, space being assumed to be homogeneous, it is also important to highlight that Eq. $(\ref{Cons1})$ is valid for a path in the $x$ direction that would start at a different location. For example, we would also have
\begin{equation}
\int_x \delta g_{xx}(y)dx = \int_x \delta g_{xx}(y+\delta y)dx = 0\,,
\end{equation}
where for convenience we did only specify the dependence of $\delta g_{xx}$ on $y$, its dependence on $x$, $z$ and $t$ being implicitly assumed. Using once again a Taylor approximation to decompose $\delta g_{xx}(y+\delta y)$, we show in a similar way as above that
\begin{equation}
\int_x \delta g_{xx,y}dx = 0\,.
\end{equation}
Still using the same approach, we show that
\begin{equation}
\int_x \delta g_{xx,x}dx = \int_x \delta g_{xx,z}dx = 0\,.
\end{equation}

Since space is assumed to be isotropic, all relations established above for a path integration along the $x$ direction are valid for paths integrations in the $y$ and $z$ directions also.

As explained above, from Eq.\ $(\ref{amoy})$ we deduce the scale factor $a$, which hence completely determines the FLRW metric tensor. Strictly speaking, the FLRW metric is defined by two parameters, namely the scale factor $a(t)$, and the value of the component $\overline{g}_{tt}$, but in practice, this latter one is constrained to be $\overline{g}_{tt} = -1$, and this is done at the expense of a modification of the time scale. The second parameter is hence replaced by the function defining this time scale modification. After having performed this time scaling, we expect that, similarly as for the spatial path, the integration of the interval along a time path calculated from the real metric converges to the one calculated from the FLRW metric. Let us hence consider a path that follows the $t$-axis ($dx = dy = dz = 0$), and let us integrate Eq.\ $(\ref{79})$ along that path. This leads to 
\begin{equation}\label{90}
\int_t ds = \int_t \sqrt{\overline{g}_{tt} + \delta g_{tt}}dt\,,
\end{equation}
where $\int_t$ means that the integration is performed along the $t$-axis. On large scales, we thus expect that 
\begin{equation}\label{91}
\int_t \sqrt{\overline{g}_{tt} + \delta g_{tt}}dt \longrightarrow \int_t \sqrt{\overline{g}_{tt}}dt\,.
\end{equation}
Following a similar approach as for a spatial path, we deduce that
\begin{equation}\label{Cons2}
\int_t \delta g_{tt}dt = 0\,.
\end{equation}

%--------------------------------------------------------------------

\section{Evolution laws for the scale factor of the representative FLRW metric}\label{S4}

At large scales, we generally assume that space presents a maximal symmetry, leading us to expect that the universe's metric globally corresponds to the FLRW metric. This approximation is motivated by the fact that when analyzing the global evolution of the universe, we only want to consider the behavior of the metric tensor at large scales, from which all local perturbations that could exist are ignored. At large scales, we also consider that it makes sense to define some global stress-energy tensor. Standard cosmological models then assume that the FLRW metric, its derived tensors and the global stress-energy tensor are simply related by the Einstein equation of general relativity, i.e., Eq.\ $(\ref{ttt})$ (or Eq.\ $(\ref{Einsteinequation})$ if we admit the existence of a non-zero cosmological constant), allowing to deduce a simple evolution law for the scale factor. This assumption should however be considered cautiously, because the metric and all related tensors as well as the stress-energy tensor are defined locally, at each point, but not globally. It is in fact not straightforward that Eq.\ $(\ref{ttt})$ is generally valid on a global scale, regardless of the fitting process we used. Remembering the simple example considered in the introduction, we should on the contrary expect that the evolution law of the scale factor depends on how the FLRW metric has been fitted on the real metric. It is thus of paramount importance to clearly derive, in function of the fitting process, this evolution law of the scale factor.

To derive this evolution law, we will proceed in two steps. In a first step, we will define the averaging process for the Einstein equation. In a second step, from the averaged Einstein equation, we will determine the evolution laws of the scale factor for the fitting process that has been presented in section $\ref{S3}$. 

\subsection{Step 1}

Averaging the Einstein equation of general relativity requires to integrate it in some way over some sub-manifold of space-time. As pointed out by several authors (see for example \cite{Buchert}), integrations over curved manifolds are in general well-defined for scalars only. Integrating tensors is indeed a questionable operation. Although we could argue that the Einstein equation itself could be integrated without worrying about issues related to the integration of tensors (see discussion further), we will proceed in a different way, by performing scalar integrations only. 

Given the FLRW metric, we note that together with the Einstein equation, we can produce two independent scalar relations. A first possibility consists in multiplying Eq.\ $(\ref{ttt})$ by $\overline{g}^{\mu\nu}$:
\begin{equation}\label{qs}
G_{\mu\nu}\overline{g}^{\mu\nu} = 8\pi G T_{\mu\nu}\overline{g}^{\mu\nu}\,.
\end{equation}
We may integrate this scalar relation over space:
\begin{equation}\label{dget}
\int_V G_{\mu\nu}\overline{g}^{\mu\nu}\sqrt{|g_{ij}|}dV = \int_V 8\pi G T_{\mu\nu}\overline{g}^{\mu\nu}\sqrt{|g_{ij}|}dV\,,
\end{equation}
where $|g_{ij}|$ is the magnitude of the determinant of the spatial part of $g_{\mu\nu}$, $\int_V$ means an integration over a spatial volume, and where $dV = dxdydz$. Once again, for convenience, the limits of the integrations are not identified, but it is assumed that the domain over which the integration is performed is sufficiently large such that perturbations are averaged over. Assuming that space is globally isotropic, we admit that on average the different spatial diagonal components of $G_{\mu\nu}$ are identical. Developing Eq.\ $(\ref{dget})$, we then get
\begin{multline}\label{dget3}
\int_V\left(\frac{3}{a^2}G_{xx}-G_{tt}\right)\sqrt{|g_{ij}|}dV =\\ \int_V 8\pi G \left(\frac{3}{a^2}T_{xx} - T_{tt}\right)\sqrt{|g_{ij}|}dV\,.
\end{multline}

A second possibility to obtain a scalar relation from the Einstein equation of general relativity and from the FLRW metric consists in multiplying Eq.\ $(\ref{ttt})$ by $\overline{g}^{\mu\alpha}\overline{g}^{\nu\beta}\overline{R}_{\alpha\beta}$. Integrating this scalar relation over space leads to:
\begin{multline}\label{dget2}
\int_V G_{\mu\nu}\overline{g}^{\mu\alpha}\overline{g}^{\nu\beta}\overline{R}_{\alpha\beta}\sqrt{|g_{ij}|}dV =\\ \int_V 8\pi G T_{\mu\nu}\overline{g}^{\mu\alpha}\overline{g}^{\nu\beta}\overline{R}_{\alpha\beta}\sqrt{|g_{ij}|}dV\,.
\end{multline}
Developing this equation gives
\begin{multline}\label{dget4}
\int_V\left(3\left(\frac{\ddot{a}}{a^3} + 2\frac{\dot{a}^2}{a^2}\right)G_{xx} - 3\frac{\ddot{a}}{a}G_{tt}\right)\sqrt{|g_{ij}|}dV = \\
\int_V 8\pi G \left[3\left(\frac{\ddot{a}}{a^3} + 2\frac{\dot{a}^2}{a^2}\right)T_{xx} -3\frac{\ddot{a}}{a}T_{tt} \right]\sqrt{|g_{ij}|}dV\,.
\end{multline}
Since the scale factor does not depend on the spatial coordinates, it can be considered as a constant in the integrations. We then multiply Eq.\ $(\ref{dget3})$ by $3\ddot{a}/a$ and subtract Eq.\ $(\ref{dget4})$ from it:
\begin{multline}
\int_V6\left(\frac{\ddot{a}}{a^3} - \frac{\dot{a}^2}{a^2}\right)G_{xx}\sqrt{|g_{ij}|}dV =\\ \int_V 6\left(\frac{\ddot{a}}{a^3} - \frac{\dot{a}^2}{a^2}\right)8\pi G T_{xx}\sqrt{|g_{ij}|}dV\,,
\end{multline}
which can be simplified:
\begin{equation}\label{tpt1x}
\int_V G_{xx}\sqrt{|g_{ij}|}dV = \int_V 8\pi G T_{xx}\sqrt{|g_{ij}|}dV\,.
\end{equation}
Using then this latter relation together with Eq.\ $(\ref{dget3})$, we deduce that
\begin{equation}\label{tpt2x}
\int_VG_{tt}\sqrt{|g_{ij}|}dV = \int_V 8\pi G T_{tt}\sqrt{|g_{ij}|}dV\,.
\end{equation}
The Eq.\ $(\ref{tpt1x})$ and $(\ref{tpt2x})$ represent the averaged Einstein equation for general relativity we propose to use.

It is important to stress that if Eq.\ $(\ref{tpt1x})$ and $(\ref{tpt2x})$ could give the impression that we integrate tensor components over space, this is in fact not the case: they are truly scalar relations, as can be seen from the way we derived them. Their scalar nature is hidden by the simplifications we made. So, these relations are well defined and completely frame invariant.

Writing finally $G_{\mu\nu}$ in function of its average part and its perturbation part, we get
\begin{equation}\label{tpt1}
\int_V\left(\overline{G}_{xx} + \delta G_{xx}\right)\sqrt{|g_{ij}|}dV = \int_V 8\pi G T_{xx}\sqrt{|g_{ij}|}dV
\end{equation}
and
\begin{equation}\label{tpt2}
\int_V\left(\overline{G}_{tt}+\delta G_{tt}\right)\sqrt{|g_{ij}|}dV = \int_V 8\pi G T_{tt}\sqrt{|g_{ij}|}dV\,.
\end{equation}

In fact, the averaged Einstein equation could have been obtained more directly. As pointed out by \cite{Hoogen}, an averaging process involves the integration (i.e., the summation) of tensors located at different points. Adding a tensor located at point $x$ to another tensor located at point $x'$ requires to parallel transport one of them to the other one along some curve. Unfortunately, in general, the value of the tensor after being parallel transported is dependent upon the selected curve, meaning that the result of the summation is not well defined. We could thus fear conceptual issues when integrating $G_{\mu\nu}$ in the left hand side and $T_{\mu\nu}$ in the right hand side of the Einstein equation. There is however one exception to this: if the tensor is null, parallel transporting it from one point to another does not depend on the curve: it remains always zero. This suggest to define the averaging process by integrating only null tensors. Hence, by defining
\begin{equation}
A_{\mu\nu} = G_{\mu\nu} - 8\pi G T_{\mu\nu}\,,
\end{equation}
which is a null tensor, we may integrate it over space without worrying on parallel transport issues, and thus
\begin{equation}
\int_V A_{\mu\nu} \sqrt{|g_{ij}|}dV = \int_V \left(G_{\mu\nu} - 8\pi G T_{\mu\nu}\right)\sqrt{|g_{ij}|}dV = 0
\end{equation}
is a frame invariant, well defined expression.

\subsection{Step 2}

We will now use the relations derived in section $\ref{S3}$ to simplify as much as possible Eq.\ $(\ref{tpt1})$ and $(\ref{tpt2})$, and deduce the evolution law of the scale factor for the considered fitting process.

First, using Eq.\ $(\ref{deltaR})$ and $(\ref{deltaG})$ to further develop Eq.\ $(\ref{tpt1})$ and $(\ref{tpt2})$, we get
\begin{multline}\label{I1}
\int_V\left(-2a\ddot{a} - \dot{a}^2 - \frac{1}{2}\delta R_{xx} + \frac{a^2}{2}\delta R_{tt} - \frac{3}{2}a\ddot{a}\delta g_{tt}\right.\\\left. - \frac{3}{2}\frac{\ddot{a}}{a}\delta g_{xx}\right)\sqrt{|g_{ij}|}dV = \int_V 8\pi G T_{xx}\sqrt{|g_{ij}|}dV\,,
\end{multline}
and
\begin{multline}\label{I2}
\int_V\left(3\frac{\dot{a}^2}{a^2} + \frac{3}{2a^2}\delta R_{xx} + \frac{1}{2}\delta R_{tt} - \frac{3}{2}\frac{2\dot{a}^2 + a\ddot{a}}{a^2}\delta g_{tt}\right.\\\left. - \frac{3}{2}\frac{2\dot{a}^2 + a\ddot{a}}{a^4}\delta g_{xx}\right)\sqrt{|g_{ij}|}dV = \int_V 8\pi G T_{tt}\sqrt{|g_{ij}|}dV\,.
\end{multline}

Concerning the determinant $|g_{ij}|$, neglecting high order terms, and admitting that on average over space $\delta g_{xx} = \delta g_{yy} = \delta g_{zz}$, we have
\begin{equation}
|g_{ij}|= a^6 + 3a^4\delta g_{xx}\,.
\end{equation}
Using a Taylor approximation, we then deduce that
\begin{equation}\label{vbp}
\sqrt{|g_{ij}|} = a^3 + \frac{3}{2}a\delta g_{xx}\,.
\end{equation}
However, according to Eq.\ $(\ref{Cons1})$, the integration of terms involving $\delta g_{xx}$ cancel (provided that the terms multiplying $\delta g_{xx}$ in the integration do not depend on $x$, which is the case here), and so the second term in the right hand side of Eq.\ $(\ref{vbp})$ may be ignored.

We thus still have to calculate $\int_V \delta R_{xx} dV$ and $\int_V \delta R_{tt} dV$. According to Eq.\ $(\ref{erpao})$, and neglecting second order terms, we have
\begin{eqnarray}\label{tyui}
&&\int_V \delta R_{xx}dV = \int_V \left(\delta \Gamma^\mu_{\ xx,\mu} - \delta \Gamma^\mu_{\ x\mu,x} + \overline{\Gamma}^\mu_{\ xx}\delta \Gamma^\nu_{\ \mu\nu}\right.\nonumber
\\
&&\left. + \overline{\Gamma}^\nu_{\ \mu\nu}\delta \Gamma^\mu_{\ xx} - \overline{\Gamma}^\nu_{\ x\mu}\delta \Gamma^\mu_{\ \nu x} - \overline{\Gamma}^\mu_{\ \nu x}\delta \Gamma^\nu_{\ x\mu}\right)dV\,.
\end{eqnarray}
By using Eq.\ $(\ref{hihi})$ and $(\ref{hoho})$, as well as the relations derived in section $\ref{S3}$, the different terms appearing in Eq.\ $(\ref{tyui})$ can then be further decomposed in function of $\overline{g}_{\mu\nu}$, $\delta g_{\mu\nu}$, and their respective derivatives. To illustrate how we can easily simplify these terms, let us consider for example the expression
\begin{equation}
\int_V \frac{1}{2}\overline{g}^{\sigma\mu}\overline{g}^{\nu\rho}\overline{g}_{\rho\beta,\alpha}\delta g_{\sigma\nu,\mu} dV\,.
\end{equation}
Since spatial derivatives vanish, this means that necessarily $\mu = t$. Since $\mu = t$, it is necessary that $\sigma = t$ also, otherwise we would have $\overline{g}^{\sigma\mu} = 0$. Moreover, we must have $\rho = \beta$ and $\alpha = t$, otherwise we would have $\overline{g}_{\rho\beta,\alpha} = 0$. We finally deduce that $\nu = \beta$, otherwise we would have $\overline{g}^{\nu\rho} = 0$. 
Performing this calculation for all terms appearing in Eq.\ $(\ref{tyui})$, we get
\begin{eqnarray}
%&&\int_V \delta \Gamma^\mu_{\ xx,\mu}dV = \int_V \left[a\dot{a}\delta g_{tt,t} + \left(\dot{a}^2+a\ddot{a}\right)\delta g_{tt}\right]dV\,,
&&\int_V \delta \Gamma^\mu_{\ xx,\mu}dV = \int_Va\dot{a}\delta g_{tt,t}dV\nonumber
\\
&& \qquad\qquad\qquad + \int_V\left(\dot{a}^2+a\ddot{a}\right)\delta g_{tt}dV\,,
\\
&&\int_V \delta \Gamma^\mu_{\ x\mu,x}dV = 0\,,
\\
&&\int_V \overline{\Gamma}^\mu_{\ xx}\delta \Gamma^\nu_{\ \mu\nu}dV = -\int_V \frac{1}{2}a\dot{a}\delta g_{tt,t}dV\,,
\\
&&\int_V \overline{\Gamma}^\nu_{\ \mu\nu}\delta \Gamma^\mu_{\ xx}dV = \int_V 3\dot{a}^2\delta g_{tt}dV\,,
\\
&&\int_V \overline{\Gamma}^\nu_{\ x\mu}\delta \Gamma^\mu_{\ \nu x}dV = \int_V \dot{a}^2\delta g_{tt}dV\,,
\\
&&\int_V \overline{\Gamma}^\mu_{\ \nu x}\delta \Gamma^\nu_{\ x\mu}dV = \int_V \dot{a}^2\delta g_{tt}dV\,.
\end{eqnarray}
We hence find that
\begin{equation}\label{Res1}
\int_V \delta R_{xx}dV = \int_V \left(\frac{1}{2}a\dot{a}\delta g_{tt,t} + \left(2\dot{a}^2 + a\ddot{a}\right)\delta g_{tt}\right)dV\,.
\end{equation}

Next, still according to Eq.\ $(\ref{erpao})$:
\begin{eqnarray}\label{tyui2}
&&\int_V \delta R_{tt}dV = \int_V \left(\delta \Gamma^\mu_{\ tt,\mu} - \delta \Gamma^\mu_{\ t\mu,t} + \overline{\Gamma}^\mu_{\ tt}\delta \Gamma^\nu_{\ \mu\nu}\right.\nonumber
\\
&&\left. + \overline{\Gamma}^\nu_{\ \mu\nu}\delta \Gamma^\mu_{\ tt} - \overline{\Gamma}^\nu_{\ t\mu}\delta \Gamma^\mu_{\ \nu t} - \overline{\Gamma}^\mu_{\ \nu t}\delta \Gamma^\nu_{\ t\mu}\right)dV\,.
\end{eqnarray}
Proceeding in the same manner as above, we get
\begin{eqnarray}
&&\int_V \delta \Gamma^\mu_{\ tt,\mu}dV = -\int_V \frac{1}{2}\delta g_{tt,tt}dV\,,
\\
&&\int_V \delta \Gamma^\mu_{\ t\mu,t}dV = -\int_V \frac{1}{2}\delta g_{tt,tt}dV\,,
\\
&&\int_V \overline{\Gamma}^\mu_{\ tt}\delta \Gamma^\nu_{\ \mu\nu}dV = 0\,,
\\
&&\int_V \overline{\Gamma}^\nu_{\ \mu\nu}\delta \Gamma^\mu_{\ tt}dV = -\int_V \frac{3}{2}\frac{\dot{a}}{a}\delta g_{tt,t}dV\,,
\\
&&\int_V \overline{\Gamma}^\nu_{\ t\mu}\delta \Gamma^\mu_{\ \nu t}dV = 0\,,
\\
&&\int_V \overline{\Gamma}^\mu_{\ \nu t}\delta \Gamma^\nu_{\ t\mu}dV = 0\,.
\end{eqnarray}
Hence
\begin{equation}\label{Res2}
\int_V \delta R_{tt}dV = -\int_V \frac{3}{2}\frac{\dot{a}}{a}\delta g_{tt,t}dV\,.
\end{equation}

Using Eq.\ $(\ref{Res1})$ and $(\ref{Res2})$, replacing $\sqrt{|g_{ij}|}$, neglecting second order terms, and reminding that the integration of terms involving $\delta g_{xx}$ cancel according to $(\ref{Cons1})$, Eq.\ $(\ref{I1})$ and $(\ref{I2})$ become, respectively,
\begin{eqnarray}\label{P1}
&&\int_V\left[-2a\ddot{a} - \dot{a}^2 - a\dot{a}\delta g_{tt,t} - \left(\dot{a}^2 + 2a\ddot{a}\right)\delta g_{tt}\right]a^3dV\nonumber
\\
&& = \int_V 8\pi G T_{xx}\sqrt{|g_{ij}|}dV\,,
\end{eqnarray}
and
\begin{equation}\label{P2}
\int_V\left(3\frac{\dot{a}^2}{a^2}\right)a^3dV = \int_V 8\pi G T_{tt}\sqrt{|g_{ij}|}dV\,.
\end{equation}
We now define the following quantities:
\begin{eqnarray}\label{b1}
\bar{\rho} &=& \frac{\int_V T_{tt}\sqrt{|g_{ij}|}dV}{\int_Va^3dV}\,,
\\\label{b2}
\bar{p} &=& \frac{1}{a^2}\frac{\int_V T_{xx}\sqrt{|g_{ij}|}dV}{\int_Va^3dV}\,,
\\
D &=& \frac{1}{a^2}\frac{\int_V \left(a\dot{a}\delta g_{tt,t} + \left(\dot{a}^2 + 2a\ddot{a}\right)\delta g_{tt}\right)a^3dV}{\int_Va^3dV}\,.
\end{eqnarray}
All these quantities are time dependent, but do not depend on the spatial coordinates. In fact, $\bar{\rho}$ simply corresponds to the spatial average of the first diagonal component of the stress-energy tensor, while $a^2\bar{p}$ corresponds to the spatial average of the other diagonal components. These definitions imply that we may write
\begin{eqnarray}\label{g1}
T_{tt}(t,x,y,z) = \bar{\rho}(t) + \delta \rho(t,x,y,z)\,,
\\\label{g2}
T_{xx}(t,x,y,z) = \bar{p}(t) + \delta p(t,x,y,z)\,,
\end{eqnarray}
where $\delta \rho$ and $\delta p$ are perturbation terms such that
\begin{equation}\label{b6}
\int_V \delta \rho \sqrt{|g_{ij}|}dV = \int_V \delta p \sqrt{|g_{ij}|}dV = 0\,.
\end{equation}

Using the above definitions, Eq.\ $(\ref{P1})$ and $(\ref{P2})$ become
\begin{eqnarray}
-2a\ddot{a} - \dot{a}^2 - Da^2 &=& 8\pi G\bar{p}a^2\,,
\\
3\frac{\dot{a}^2}{a^2} &=& 8\pi G\bar{\rho}\,.
\end{eqnarray}
Combining these equations adequately, we get
\begin{eqnarray}\label{F1}
\frac{\ddot{a}}{a} &=& -\frac{4\pi G}{3}\left(\bar{\rho}+3\bar{p}\right) - \frac{D}{2}\,,
\\\label{F2}
\frac{\dot{a}^2}{a^2} &=& \frac{8\pi G}{3}\bar{\rho}\,.
\end{eqnarray}	
We notice that Eq.\ $(\ref{F2})$ is identical to the first Friedmann equation. This is an important result that confirms that using the Einstein equation to study the global behavior of space-time is allowed if the FLRW metric has been fitted according to the constraints derived in section $\ref{S3}$.

On the other hand, Eq.\ $(\ref{F1})$ differs from the second Friedmann equation due to the presence of $D$. This equation provides information on the average perturbation $\delta g_{tt}$, but it can be presented in a more useful manner. We will first simplify Eq.\ $(\ref{F1})$. Therefore, we differentiate Eq.\ $(\ref{F2})$ with respect to time:
\begin{equation}\label{b4}
\frac{\ddot{a}}{a} = \frac{8\pi G}{3}\bar{\rho} + \frac{4\pi G}{3}\frac{a}{\dot{a}}\dot{\bar{\rho}}\,.
\end{equation}
To further simplify this expression, we will make use of the conservation equation:
\begin{eqnarray}\label{b8}
0 &=& g^{\sigma\mu}\nabla_\sigma T_{\mu \nu}\nonumber
\\
&=& g^{\sigma\mu}\left(\partial_\sigma T_{\mu \nu} - \Gamma^\lambda_{\ \sigma\mu}T_{\lambda \nu} - \Gamma^\lambda_{\ \sigma \nu}T_{\mu \lambda}\right)\,.
\end{eqnarray}
This is a vectorial relation of the form $B_\nu = 0$. As previously, together with the FLRW metric, we could make from this two independent scalar relations, by multiplying it first by $\overline{g}^{\nu\xi}B_\xi$, then by $\overline{g}^{\nu\alpha}\overline{g}^{\xi\beta}\overline{R}_{\alpha\beta}B_\xi$. These scalar relations could be adequately combined, and then integrated to obtain a useful relation. However, as argued above, since $B_\nu$ is a null tensor, its integration over space is well defined. We can hence proceed in a much easier way (but the same result would be obtained by the first method).

So, integrating the $t$ component of Eq.\ $(\ref{b8})$ over space, developing the terms in function of their average and perturbation parts and dividing by $\int_V \sqrt{|g_{ij}|}dV$, we obtain
\begin{equation}\label{b3}
0 = \frac{\int_V B_t \sqrt{|g_{ij}|}dV}{\int_V \sqrt{|g_{ij}|}dV}\,,
\end{equation}
where
\begin{multline}
B_t = \left(\overline{g}^{\sigma\mu} + \delta g^{\sigma\mu}\right)\left[\partial_\sigma T_{\mu t} - \left(\overline{\Gamma}^\lambda_{\ \sigma\mu} + \delta \Gamma^\lambda_{\ \sigma\mu}\right)T_{\lambda t}\right.\\\left. - \left(\overline{\Gamma}^\lambda_{\ \sigma t} + \delta \Gamma^\lambda_{\ \sigma t}\right)T_{\mu \lambda}\right]\,.
\end{multline}
Proceeding as before, by neglecting higher order terms, by taking into account the properties expressed by Eq.\ $(\ref{Cons1})$ and $(\ref{b6})$, and using the definitions $(\ref{b1})$ and $(\ref{b2})$, we can simplify Eq.\ $(\ref{b3})$ as
\begin{equation}
\dot{\bar{\rho}} = -3\frac{\dot{a}}{a}\left(\bar{\rho} + \bar{p}\right)\,.
\end{equation}
Using then this last relation into Eq.\ $(\ref{b4})$, we get
\begin{equation}\label{b5}
\frac{\ddot{a}}{a} = -\frac{4\pi G}{3}\left(\bar{\rho}+3\bar{p}\right)\,.
\end{equation}
Comparing finally Eq.\ $(\ref{b5})$ with Eq.\ $(\ref{F1})$ we deduce that $D = 0$. This provides information on the evolution of the perturbation $\delta g_{tt}$. Let us investigate this.

We notice that $D$ can also be written as
\begin{eqnarray}\label{odi}
D &=& \frac{1}{a^2\dot{a}}\frac{\int_V \frac{\partial}{\partial t}\left(a\dot{a}^2\delta g_{tt}\right)dV}{\int_V a^3dV}\nonumber
\\
&=& \frac{8\pi G}{3a^2\dot{a}}\frac{\frac{\partial}{\partial t}\left(\int_V a^3\bar{\rho}\delta g_{tt}dV\right)}{\int_V a^3dV}\,,
\end{eqnarray}
where we made use of Eq.\ $(\ref{F2})$. The fact that $D = 0$ implies that the derivative in Eq.\ $(\ref{odi})$ cancels, and hence that the expression that is derived is equal to some constant. Obviously, this constant is proportional to the volume over which the integration is performed. We thus have
\begin{equation}\label{fff1}
\int_V a^3\bar{\rho} \delta g_{tt}dV = \alpha \int_V dV\,,
\end{equation}
where $\alpha$ is a constant. Since $a$ and $\bar{\rho}$ do not depend on the spatial coordinates, we deduce
\begin{equation}\label{fff}
\frac{\int_V \delta g_{tt}dV}{\int_V dV} = \frac{\alpha}{a^3\bar{\rho}}\,.
\end{equation}
The left hand side represents the spatial average of $\delta g_{tt}$.

Let us show that for a representative FLRW metric we necessarily have $\alpha = 0$. We therefore start from Eq.\ $(\ref{Cons2})$ and integrate it over space:
\begin{equation}
\int_V \left(\int_t \delta g_{tt} dt\right)dV = 0\,,
\end{equation}
which can be written as
\begin{equation}
\int_t \left(\int_V \delta g_{tt} dV\right)dt = \int_t \left(\frac{\alpha}{a^3\bar{\rho}} \int_VdV\right) dt = 0\,.
\end{equation}
Since $\rho$ and $a$ are positive variables, the previous equation will be verified only if $\alpha = 0$. We thus conclude that a representative FLRW metric is also characterized by the fact that
\begin{equation}\label{vvv}
\int_V \delta g_{tt}dV = 0\,.
\end{equation}

%--------------------------------------------------------------------

\section{Redshift and luminosity distance measurements}\label{S5}

We verify in this section that the interpretation of the measurements performed to determine the evolution of the scale factor implicitly lead to a fitting of the FLRW metric according to the constraints identified in section $\ref{S3}$. Such measurements consist in redshift and luminosity distance measurements.	

\subsection{Redshift measurements}

Let us consider a source emitting light with a known temporal characteristic. A first signal is emitted at time $t_1$ by such a source located at $x = x_1$ and reaches at time $t_2$ an observer located at $x = x_2$. A second signal is emitted from the same source at time $t_1 + \Delta t_1$ and reaches the observer at time $t_2+\Delta t_2$.

Light follows a null geodesic. Integrating Eq.\ $(\ref{79})$ along such a geodesic by considering the real metric yields for the first signal
\begin{equation}
\int_{t_1}^{t_2} \sqrt{\frac{-g_{tt}}{g_{xx}}}dt = \int_{x_1}^{x_2}dx\,.
\end{equation}
Considering the equivalent relation for the second signal we show that
\begin{equation}\label{ggqq}
\int_{t_1}^{t_1+\Delta t_1}\sqrt{\frac{-g_{tt}}{g_{xx}}}dt = \int_{t_2}^{t_2+\Delta t_2}\sqrt{\frac{-g_{tt}}{g_{xx}}}dt\,.
\end{equation}
For a small variation in time, the components of the metric may be considered as constant, and we deduce that
\begin{equation}\label{ppp}
\sqrt{\frac{-g_{tt}(x_1,t_1)}{g_{xx}(x_1,t_1)}}\Delta t_1 = \sqrt{\frac{-g_{tt}(x_2,t_2)}{g_{xx}(x_2,t_2)}}\Delta t_2\,.
\end{equation}
We define $\lambda_1 = \sqrt{-g_{tt}(x_1,t_1)}\Delta t_1$ and $\lambda_2 = \sqrt{-g_{tt}(x_2,t_2)}\Delta t_2$. These quantities represent the proper time between two signals as measured at the source and the observer, respectively. Then Eq.\ $(\ref{ppp})$ becomes
\begin{equation}\label{ivc}
\sqrt{a^2(t_1) + \delta g_{xx}(x_1,t_1)} = \frac{\lambda_1}{\lambda_2}\sqrt{a^2(t_2) + \delta g_{xx}(x_2,t_2)}\,.
\end{equation}
This equation is the one that should be used to determine the scale factor from measurements. However, in practice, we use the following equation:
\begin{equation}\label{pep}
a(t_1) = \frac{\lambda_1}{\lambda_2}a(t_2)\,.
\end{equation}
By imposing that Eq.\ $(\ref{pep})$ is equivalent to Eq.\ $(\ref{ivc})$ we implicitly impose some constraint, that contributes to defining the fitting process. Let us first consider the left hand side of Eq.\ $(\ref{ivc})$. Obviously, at the source of the signals (position $x_1$), $\delta g_{xx}$ could be everything, meaning that we cannot know its value for a specific source. But if redshift measurements are carried out for several sources occurring at the same temporal variable $t$, statistics apply. If we thus assume that on average over space $\delta g_{xx}=0$, the left hand side of Eq.\ $(\ref{ivc})$ reduces to the one of Eq.\ $(\ref{pep})$. This assumption exactly corresponds to the constraint expressed by Eq.\ $(\ref{Cons1})$. This means that the way we interpret redshift measurements is compatible with the expectations of a representative FLRW metric.

On the other hand, since measurements are performed by the observer at one single position, $\delta g_{xx}(x_2,t_2)$ has a fixed value which cannot be ignored in general. The correction related to $\delta g_{xx}(x_2,t_2)$ would obviously slightly modify the results for $a(t_1)$ as obtained by the usual practice, but it would not alter the global trend of its evolution, except maybe for SNIa in the local void region around our galaxy, see \cite{Alexander} for example, meaning that it could not explain the observed accelerated expansion (except if the aforementioned local void region is extremely large).

\subsection{Luminosity distance measurements}

The luminosity distance $d_L$ is defined as
\begin{equation}\label{qk}
d_L^2 = \frac{L}{4\pi F}\,,
\end{equation}
where $L$ is the absolute luminosity of the source (supposed to be known) and $F$ is the flux measured by the observer. As for the redshift measurements, we should take into account local perturbations at the position of the observer to correctly determine the flux. But again, this would constitute a constant correction only, and this would not alter the global trend of the evolution of the scale factor. Also, the absolute luminosity represents an amount of energy per unit time. This parameter is expressed in function of the proper time of the source. So, in theory, we should take into account the local perturbation $\delta g_{tt}$ to correctly express $L$ in Eq.\ $(\ref{qk})$. But here also, when performing luminosity distance measurements on a large sample, statistics apply, and if we assume that on average over space $\delta g_{tt} = 0$, the correction that has to be applied on $L$ cancels. It happens that this assumption is also verified by a representative FLRW metric, see Eq.\ $(\ref{vvv})$. We hence conclude that Eq.\ $(\ref{qk})$ together with Eq.\ $(\ref{pep})$ are coherent with the constraints expected for a representative FLRW metric.

%--------------------------------------------------------------------

\section{Discussion}\label{S6}

On a theoretical point of view, from our investigation, it appears that the way measurements are interpreted is appropriate to determine the FLRW metric that correctly approximates the real one. Indeed, the as fitted FLRW metric would be such that on large scales, spatial dimensions and time intervals would be equivalent to the respective ones determined from the real metric.	

On a practical point of view, however, this conclusion could potentially be invalidated for the following reason: for the redshift measurements, Eq.\ $(\ref{pep})$ was shown to be equivalent to Eq.\ $(\ref{ivc})$ by imposing that on average over space $\delta g_{xx}=0$ at the source. Similarly, for the luminosity measurements, Eq.\ $(\ref{qk})$ was obtained by imposing that on average over space $\delta g_{tt}=0$ at the source. Both conditions would indeed be verified if measurements were carried out on sources being equally distributed over space. But are they? 

We should remind that the fact that space is not perfectly homogeneous is precisely due to an inhomogeneous distribution of matter. Matter is mainly concentrating in overdense regions, leaving other underdense regions of almost void. Statistically, we would expect that most of the SNIa happen where matter is present, hence in the overdense regions. The more matter is present, the more stars we will have, and the more chances we have to observe a SNIa. We could thus fear that most of the SNIa that have been observed are located in overdense regions. If this was the case, that would mean that we are fitting the FLRW metric in a biased manner, by considering only information coming from these overdense regions, and excluding the one from the underdense regions.

What does this mean in practice? Let us consider once again the redshift measurements. As evidenced by Eq.\ $(\ref{ivc})$, the inhomogeneous property of space implies that measurements carried out on different SNIa occurring at time $t$ would lead to slightly different values of the scale factor. Different values are indeed observed. Some part of these differences is certainly related to the measurement uncertainties, but another part, even small, is related to the inhomogeneities. At the end, the value that we attribute to $a$ for time $t$ corresponds to the average of all measurements carried out for that time, written as $\langle a \rangle$. Since SNIa are more likely to occur in overdense regions, this means that the average is performed by using a weight factor corresponding to the density, such that more importance is given to regions having a large density:
\begin{eqnarray}
\langle a \rangle &=& \frac{\int_V \rho \left(\sqrt{a^2 + \delta g_{xx}}\right)\sqrt{|g_{ij}|}dV}{\int_V\rho \sqrt{|g_{ij}|}dV}\nonumber
\\
&\simeq& a + \frac{a^2}{2}\frac{\int_V \rho \delta g_{xx}dV}{\int_V\rho \sqrt{|g_{ij}|}dV}\,.
\end{eqnarray}
But since in practice we use Eq.\ $(\ref{pep})$, we implicitly assume that $\langle a \rangle = a$, and hence that
\begin{equation}\label{v1}
\int_V \rho \delta g_{xx}dV = 0\,.
\end{equation}
This last equation differs from the constraint expressed by Eq.\ $(\ref{Cons1})$, and this means that we are fitting the FLRW metric in a different way.

In a similar way, considering how luminosity distance measurements are performed, we deduce that we also assume that
\begin{equation}\label{v2}
\int_V \rho \delta g_{tt}dV = 0\,.
\end{equation}
Again, this last equation differs from the constraint expressed by Eq.\ $(\ref{vvv})$.

For the representative FLRW metric, we have shown above that the Friedmann equation was indeed applicable. Now, since we are fitting the FLRW metric by using different constraints than the ones established for the representative FLRW metric, the Friedmann equation would probably not remain applicable as such anymore. To see how the Friedmann equation should be modified, we will proceed as we did in section $\ref{S4}$, by using the new constraints.

For the first step, we multiply Eq.\ $(\ref{qs})$ by $\overline{g}^{\sigma\lambda}T_{\sigma\lambda}$. 
\begin{equation}
G_{\mu\nu}\overline{g}^{\mu\nu}\overline{g}^{\sigma\lambda}T_{\sigma\lambda} = 8\pi G T_{\mu\nu}\overline{g}^{\mu\nu}\overline{g}^{\sigma\lambda}T_{\sigma\lambda}\,.
\end{equation}
Once again, this is a scalar relation, and we may integrate it over space. By noting that in a matter dominated universe $|\overline{g}^{\sigma\lambda}T_{\sigma\lambda}| \simeq T_{tt}$, i.e. the local density $\rho$, we then obtain
\begin{multline}
\int_V\rho\left(\frac{3}{a^2}G_{xx}-G_{tt}\right)\sqrt{|g_{ij}|}dV =\\ \int_V 8\pi G\rho \left(\frac{3}{a^2}T_{xx} - \rho\right)\sqrt{|g_{ij}|}dV\,.
\end{multline}
Multiplying then Eq.\ $(\ref{qs})$ by $\overline{g}^{\mu\alpha}\overline{g}^{\nu\beta}\overline{R}_{\alpha\beta}\overline{g}^{\sigma\lambda}T_{\sigma\lambda}$, integrating the obtained scalar relation over space, we get
\begin{multline}
\int_V\rho\left(3\left(\frac{\ddot{a}}{a^3} + 2\frac{\dot{a}^2}{a^2}\right)G_{xx} - 3\frac{\ddot{a}}{a}G_{tt}\right)\sqrt{|g_{ij}|}dV = \\
\int_V 8\pi G\rho \left[3\left(\frac{\ddot{a}}{a^3} + 2\frac{\dot{a}^2}{a^2}\right)T_{xx} -3\frac{\ddot{a}}{a}\rho \right]\sqrt{|g_{ij}|}dV\,.
\end{multline}
Combining these two last equations adequately, we show that
\begin{equation}
\int_V\rho\left(\overline{G}_{xx} + \delta G_{xx}\right)\sqrt{|g_{ij}|}dV = \int_V 8\pi G\rho T_{xx}\sqrt{|g_{ij}|}dV
\end{equation}
and
\begin{equation}\label{coc}
\int_V\rho\left(\overline{G}_{tt}+\delta G_{tt}\right)\sqrt{|g_{ij}|}dV = \int_V 8\pi G\rho^2\sqrt{|g_{ij}|}dV\,.
\end{equation}
Let us more particularly consider Eq.\ $(\ref{coc})$. Using the constraints expressed by Eq.\ $(\ref{v1})$ and $(\ref{v2})$, and proceeding in a similar way as we did in section $\ref{S4}$, we can simplify Eq.\ $(\ref{coc})$ as follows:
\begin{equation}\label{P2bis}
\int_V\rho \left(3\frac{\dot{a}^2}{a^2}\right)\sqrt{|g_{ij}|}dV = \int_V 8\pi G\rho^2\sqrt{|g_{ij}|}dV\,.
\end{equation}
Replacing $\rho$ by $\bar{\rho} + \delta\rho$ (according to Eq.\ $(\ref{g1})$) in the right hand side of this equation, we get
\begin{equation}\label{Fried}
\frac{\dot{a}^2}{a^2} = \frac{8\pi G}{3}\left(\bar{\rho} + \rho_\Lambda\right)\,,
\end{equation}
where
\begin{equation}
\rho_\Lambda = \frac{\int_V\rho\delta\rho\sqrt{|g_{ij}|}dV}{\int_V\rho\sqrt{|g_{ij}|}dV}\,.
\end{equation}
Using once again the fact that $\rho = \bar{\rho} + \delta\rho$, we show that
\begin{equation}
\int_V\rho\delta\rho\sqrt{|g_{ij}|}dV = \int_V\left(\delta\rho\right)^2\sqrt{|g_{ij}|}dV + \bar{\rho}\int_V\delta\rho\sqrt{|g_{ij}|}dV\,.
\end{equation}
By virtue of Eq.\ $(\ref{b6})$, the last term cancels, and hence
\begin{eqnarray}\label{rhol}
\rho_\Lambda &=& \frac{\int_V\left(\delta\rho\right)^2\sqrt{|g_{ij}|}dV}{\int_V\rho\sqrt{|g_{ij}|}dV}\nonumber
\\
&=& \frac{\int_V\left(\delta\rho\right)^2\sqrt{|g_{ij}|}dV}{\int_V\sqrt{|g_{ij}|}dV}\frac{\int_V\sqrt{|g_{ij}|}dV}{\int_V\rho\sqrt{|g_{ij}|}dV}\nonumber
\\
&=& \frac{\langle\left(\delta\rho\right)^2\rangle}{\bar{\rho}}\,,
\end{eqnarray}
where
\begin{equation}\label{lam}
\langle\left(\delta\rho\right)^2\rangle = \frac{\int_V\left(\delta\rho\right)^2\sqrt{|g_{ij}|}dV}{\int_V\sqrt{|g_{ij}|}dV}
\end{equation}
represents the mean squared deviation of the density with respect to its average. It is zero for a perfectly homogeneous universe, and increases as matter inhomogeneities are more pronounced. 

Hence, the new evolution law of the scale factor is described by Eq.\ $(\ref{Fried})$. It differs from the one applicable for the representative metric due to the presence of a new term, $\rho_\Lambda$, which according to Eq.\ $(\ref{rhol})$ is clearly positive. If sufficiently large, this term will imply an acceleration of the scale factor, and hence accounts for the dark energy effect. This term is however not expected to be constant in time.

In order to understand the behavior of Eq.\ $(\ref{Fried})$, let us write it as
\begin{equation}\label{ici}
\frac{\dot{a}^2}{a^2} = \frac{8\pi G}{3}\bar{\rho}\left(1 + \frac{\langle\left(\delta\rho\right)^2\rangle}{\bar{\rho}^2}\right)\,,
\end{equation}
At the early ages of the matter dominated period of the universe, the average density $\bar{\rho}$ was large, and at large scales, matter was quite smoothly distributed, meaning that $\langle\left(\delta\rho\right)^2\rangle/\bar{\rho}^2$ was small. This implies that in the parentheses of Eq.\ $(\ref{ici})$, the dominating term is 1, and the evolution law reduces to the one applicable for the representative metric. This correspond to the Friedmann equation with a null cosmological constant. But over time, while the average density decreases, large-scale structures develop, and the matter distribution presents more and more inhomogeneities, meaning that $\langle\left(\delta\rho\right)^2\rangle/\bar{\rho}^2$ increases. In the parentheses of Eq.\ $(\ref{ici})$, the second term increases progressively, and at some time, becomes the dominating one. The scale factor $a$ now presents an accelerated evolution. This behavior is in accordance with the observations. It must however be stressed that this behavior is an illusion that results from an inadequate fitting of the FLRW metric on the real metric, because of the bias of the measurements.

The fact that the accelerated expansion as observed from SNIa measurements could be an illusion has already been suggested by several authors, see for example \cite{Alexander}, \cite{Wiltshire2}, \cite{Ishak}, \cite{Colin} or \cite{Iguchi}. The explanation given in those articles is generally based on the assumption that our galaxy is located in an underdense region. The present article also suggests that the observed accelerated expansion is only apparent, but the explanation differs in the way that it is not based on a local effect, but on a global effect, related to an inadequate approximation of the real metric. 

In order to further support or disprove the proposed theory, a numerical estimation should be provided for the current value of $\rho_\Lambda$, as defined by Eq.\ $(\ref{rhol})$, and should be compared with the value deduced form the as observed evolution of the scale factor. This is however out of the scope of this article.

%--------------------------------------------------------------------

\section{Conclusion}

Starting from the considerations that space is not perfectly homogeneous and isotropic, that the FLRW metric is only an approximation of the real universe's metric, and by making an analogy with a simple example, we first stressed that the approximate FLRW metric has to be fitted on the real one, and that this fitting can be done in different ways.

By using a perturbation approach, we identified in a first step what would be an adequate fitting, and this led us to define the representative FLRW metric, i.e., a metric which has been fitted such that its evolution can be considered as representative of the real one when inhomogeneities are averaged over. This metric is characterized by the fact that on large scales the integration of the interval over some path by using the FLRW metric tends to the integration of the interval over the same path by using the real metric. From the application of this definition on specific paths, important constraints have been derived on how we should fit the FLRW metric. Using these constraints, we have then verified that the average Einstein equation in which the cosmological constant is assumed to be zero led indeed to the Friedmann equations still without any cosmological constant.

In a second step, we have shown that the way measurements on SNIa are interpreted implies the same constraints as the ones mentioned above, indicating that such measurements lead to implicitly fit the FLRW metric such that it corresponds to the representative one. However, we identified a bias in the measurements, related to the fact that SNIa are not randomly distributed over space, but are probably mostly located in regions were matter is largely present, i.e., in overdense regions.

In a third step, we showed that this bias led to fit the FLRW metric differently from the representative one, and that as a consequence, a new term appears in the Friedmann equation, equivalent to a cosmological constant, which could account for an acceleration of the scale factor of the as fitted metric. We hence showed that this apparent acceleration is an illusion that can be explained by and inadequate fitting of the FLRW metric, without needing to introduce the dark energy assumption. 

%----------------------------------------------------------------------------------------
%	REFERENCE LIST
%----------------------------------------------------------------------------------------

%\bibliography{Article1}
\bibliographystyle{unsrt}

%----------------------------------------------------------------------------------------

\end{document}